\documentstyle[12pt,epsfig]{article}
\newcommand{\be}{\begin{equation}}
\newcommand{\ee}{\end{equation}}

\newcommand{\bbf}{\bf}
\newcommand{\ssl}{\sl}

\newcommand{\T}{\mbox{\bf T}}

\newcommand{\bea}{\begin{eqnarray}}
\newcommand{\eea}{\end{eqnarray}}
\newcommand{\1}{1\!{\rm l}}

\begin{document}

\begin{titlepage}
\begin{flushright}
FSUJ-TPI 00/02 \\
hep-th/0002006  \\
\end{flushright}

\vspace{2  cm}

\begin{center}
\Large{The Tale of Gravitational Sphaleron}\footnote{
Talk presented at the International Workshops 
``Supersymmetry and Quantum\\
 Symmetries''
and ``Quantum Gravity and Strings'', Dubna 99. \\
Supported by the Deutsche
Forschungsgemeinschaft, DFG-Wi 777/4-1, 777/4-2.}

\vskip12mm
\large
Mikhail S. Volkov

\vspace{3mm}
{\small\sl
Institute for Theoretical Physics\\
Friedrich Schiller University of Jena\\
Max-Wien Platz 1, D-07743\\
Jena, Germany\\
e-mail: vol@tpi.uni-jena.de}

\end{center}

\vspace{1 cm}

{Type I string theory can be dimensionally reduced on 
group manifolds. The compactification on 
$S^3\times S^3$ leads to the N=4 gauged SU(2)$\times$SU(2)
supergravity in four dimensions, which admits the BPS monopole
type non-Abelian vacuum. The reduction on 
$S^3\times AdS_3$ gives  the  Euclidean   
N=4 gauged SU(2)$\times$SU(1,1) supergravity
admitting a globally regular supersymmetric non-Abelian 
background. The latter
can be analytically continued to the 
Lorentzian sector, which gives the  regular, unstable particle-like
configuration known as gravitational sphaleron.
When lifted to D=10, the Euclidean vacuum describes 
a deformation of the D1--D5 brane system.
}

\end{titlepage}
\newpage

\noindent
{\bf Introduction.--}
There are several reasons to 
study systems with interacting gravitational and non-Abelian 
gauge fields with the Lagrangian of the type
\be                     \label{lag}
{\cal L}=-\frac{1}{16\pi G}\, R -\frac{1}{4g^2}\,
F^a_{\mu\nu}F^{a\mu\nu}+\ldots \, 
\ee
Here dots stand for other possible fields, which can be
scalars, Abelian vectors, etc. Also
the gauge coupling constant $g$ can be position-dependent, 
which is the case when a dilaton is present. First, 
all modern models of particle physics include Yang-Mills fields as a 
basic element and, since
the energies under consideration are climbing higher and higher,
the effects of gravity should also be considered. Besides, 
as suggested by the recent brane-world scenarios, even at low 
energies
gravity might be important \cite{Randall99}. 
From the classical General Relativity point of view, the model 
(\ref{lag}) can be viewed as a natural generalization of the 
Einstein-Maxwell theory. 
This suggests studying its solutions and, 
in particular, checking  
whether the standard electrovacuum  theorems  still apply. 
Surprisingly, the answer to the latter question is negative -- 
the theory admits a large variety of new bizarre solutions, 
such as hairy black holes, say, whose existence manifestly 
contradicts the 
conventional wisdom like the no-hair conjecture \cite{Volkov98}. 
Unfortunately, most of these solutions are available only 
numerically,
which is due to the high complexity of the equations. 

Finally, string theory is 
perhaps the most natural place where 
models
of the type (\ref{lag}) apply.  The Lagrangian 
(\ref{lag}) arises
in the low energy limit of heterotic string theory, and in 
type I and II string theories and M-theory 
compactified on manifolds with non-Abelian 
isometries. 
Supersymmetric string vacua  play an important 
role in the analysis of string theory \cite{Duff95,Youm97}.
However, apart from stringy monopoles \cite{Gauntlett93} 
and related solutions \cite{Duff95}
obtained via the heterotic five-brane construction 
\cite{Strominger90}
(see also \cite{Gibbons94a,Gibbons95a}), most of the literature
is devoted to solutions with Abelian gauge fields. 
This is easily understood, since  non-Abelian solutions are 
much more 
difficult to study than those from the Abelian sector. 
On the other hand, it is to be expected that also configurations 
with
non-Abelian gauge fields will eventually play an important role.
In particular, this is suggested by the AdS/CFT correspondence,
which focuses on backgrounds of gauged supergravity models 
\cite{Maldacena99}. 

\noindent
{\bf The tale of the EYMS sphaleron.--}
As was mentioned above, only very few supergravity solutions with
non-Abelian gauge fields are known explicitly. 
Of these we would like to describe
one which is particularly 
interesting in view of the surprising developments it has caused. 
This solution was originally discovered numerically in the 
Einstein-Yang-Mills-Dilaton (EYMD) model for the gauge group 
SU(2) \cite{Donets93,Lavrelashvili93,Bizon93a}
(in which case one has in Eq.(\ref{lag}) $g={\rm e}^{-\phi}$,
and there is also the standard kinetic term for the dilaton $\phi$). 
The solution describes
a static, spherically symmetric particle-like object with finite 
ADM mass
and a globally regular geometry. The gauge field is 
essentially non-Abelian and purely magnetic 
with zero total charge -- due to the fast 
fall-off at infinity.  
It was realized that the solution is unstable 
\cite{Lavrelashvili93}, which however is not a drawback. 
On the contrary, it turns out that the solution admits an 
elegant interpretation \cite{Galtsov91a} as a sphaleron 
\cite{Manton83,Klinkhamer84}. Specifically, one can 
show that the solution relates to the top of the potential
barrier between the topological vacua in the EYMD theory. 
This suggests that it might be responsible for some kind
of fermion number non-conservation in the theory. Let us 
call this solution EYMD sphaleron. 

A number of surprising regularities in the parameters 
of the solution have been observed \cite{Lavrelashvili93}. 
Specifically, the parameters fulfill simple
algebraic relations with rational coefficients, which is
usually not expected for numerically obtained solutions. It was therefore
conjectured that the solution had  a
 hidden symmetry, which hopefully
might even help to obtain it analytically. 
A part of this symmetry was soon identified as the dilatational 
invariance \cite{Donets93b}, and this accounted for some of the
relations. However, other relations remained unexplained,
and all attempts to find the solution in a closed analytical form failed.

It was then natural to check whether supersymmetry could
account for the remaining hidden symmetry. The immediate
obstacle, however, was the instability of the solution, which 
made the existence of supersymmetry extremely unlikely. 
Nevertheless, it was still possible to look for gauged 
supergravities containing the EYMD theory or some deformation of it.
Such theories might admit supersymmetric vacua similar 
to the EYMD sphaleron solution. The appropriate model
was soon found and studied in \cite{Chamseddine97,Chamseddine98} --
it was the N=4 gauged SU(2)$\times$SU(2)  
supergravity,
also known as the Freedman--Schwarz (FS) model \cite{Freedman78}.
This model is very similar to the EYMD theory, but contains
in addition the dilaton potential
\be                                                  \label{0}
{\rm U}(\phi)=-\frac18\,(g_{(1)}^2+g_{(2)}^2)\, {\rm 
e}^{-2\phi}\, ,
\ee
where $g_{(1)}$ and $g_{(2)}$ are the gauge coupling constants
for the two non-Abelian gauge fields. 
It was shown in 
\cite{Chamseddine97,Chamseddine98} that 
the FS model admits essentially non-Abelian vacua of the 
BPS monopole type. These solutions are static, spherically 
symmetric, globally regular and have purely magnetic gauge field
with unit magnetic charge -- even though there is no Higgs field
in the theory. Next, they preserve 1/4 of the supersymmetries,
and are probably stable, but not asymptotically flat --
due to the unbounded dilaton potential.   
The discovery of these solutions has enlarged somewhat
the very narrow family of analytically known 
configurations for gravitating Yang-Mills fields. 
Moreover, it was shown in  
\cite{Chamseddine98} that the FS model itself can be obtained
via dimensional reduction of the N=1, D=10 supergravity
on $S^3\times S^3$ manifold. This established a direct link
to string theory and showed that any solution of the FS model
can be lifted to ten dimensions. 

However, the situation with the EYMD sphaleron solution 
and its remaining hidden symmetry
remained obscure. In order to obtain this solution within a gauged
supergravity model it was necessary to get rid of
the dilaton potential. However, the latter is present  practically in all 
models.
The following 
trick was therefore employed in \cite{Volkov99}: to truncate
the FS model to the purely magnetic sector and then to pass to 
 imaginary values of the gauge coupling constant $g_{(2)}$:
\be                                             \label{0a}
g_{(2)}\to i g_{(2)}.
\ee
For $|g_{(2)}|=g_{(1)}$ the potential vanishes. 
Surprisingly, such a formal trick
does not destroy supersymmetry -- the replacement 
(\ref{0a}) in the FS fermion supersymmetry transformations
leads to non-trivial Bogomol'nyi equations. These admit
asymptotically flat solutions, one of which exactly coincides
with the EYMD sphaleron !  To recapitulate, 
the EYMD sphaleron fulfills the first order 
Bogomol'nyi equations, and this explains all
the remaining regularities in the parameters of the solution. 

As the existence of Bogomol'nyi equations is usually related
to supersymmetry, it was tempting to 
conclude that the solution is supersymmetric. 
However, it was unclear how to reconcile supersymmetry
with the instability of the solution. Moreover, the 
replacement (\ref{0a}) is a rather formal trick, and it is 
unclear to what supergravity model it leads. 
The resolution of these puzzles is interesting. 
It was conjectured
in \cite{Volkov99} and verified in 
\cite{Volkov99a} that there is another, hitherto unknown  
consistent {\sl Euclidean}
gauged supergravity whose special truncation can be 
related to the purely magnetic sector of the 
FS model via the replacement (\ref{0a}). 

The new supergravity can be obtained via a consistent dimensional 
reduction of the N=1, D=10 SUGRA on $S^3\times AdS_3$. 
Since the metric on the internal 6-space is not 
positive-definite,
the timelike coordinate of the ten-dimensional space  is 
viewed as one of the internal coordinates, while
the remaining four-space becomes Euclidean. 
As a result, one obtains the  Euclidean N=4, D=4 
gauged supergravity called
Euclidean Freedman-Schwarz (EFS) model. 
The fields of the new theory are similar to those
of the FS model -- the bosonic sector contains gravitational
field $g_{\mu\nu}$, two non-Abelian gauge fields 
$A^{(1)a}_\mu$ and $A^{(2)a}_\mu$  with coupling constants 
$g_{(1)}$ and $g_{(2)}$, the axion ${\bf a}$ 
and the dilaton $\phi$. The principal new features 
of the theory is that it lives in Euclidean space, 
the gauge group is now SU(2)$\times$SU(1,1), and the dilaton
potential is obtained from (\ref{0}) via replacing the 
sum of the two gauge coupling constants by their
difference, $g_{(1)}^2-g_{(2)}^2$. The latter property
is due to the opposite signs of the curvature of 
$S^3$ and $AdS_3$.  

Now, one can consider ``static and purely magnetic'' 
(these notions do not have an invariant meaning in the Euclidean
regime) sector of the EFS model.
In this sector the SU(1,1) gauge field 
$A^{(2)a}_\mu$ is zero,  $A^{(1)a}_\mu$ is ``purely magnetic'',
and there is a hypersurface orthogonal ``timelike'' Killing
vector. Then one discovers that all bosonic equations and
fermionic SUSY transformations are related to those of the 
static and purely magnetic sector of the FS model via
the replacement (\ref{0a}). Next, one finds that the Euclidean
theory admits a globally regular non-Abelian background, 
and this can be analytically continued to the Lorentzian
sector, which gives precisely the EYMD sphaleron solution. 
This solves all puzzles:  the EYMD sphaleron
is not a supersymmetric solution. However, it becomes truly
supersymmetric upon continuation to the Euclidean sector, 
simply via replacement time by imaginary
time, $t\to it$. Both Euclidean and Lorentzian 
configurations fulfill the same system of Bogomol'nyi equations. 
The instability of the solution occurs only 
in the Lorentzian sector, where there is no
supersymmetry.

Summarizing, the analysis of the EYMD sphaleron 
has led to the disclosure of the Kaluza-Klein origin of the FS 
model, to the supersymmetric monopoles,  
and to the new Euclidean gauged supergravity 
obtained from type I string theory via the dimensional reduction.  
The sphaleron can be analytically continued to the Euclidean sector
to become the supersymmetric vacuum of the new supergravity. 
These features are
briefly summarized in the following diagram.


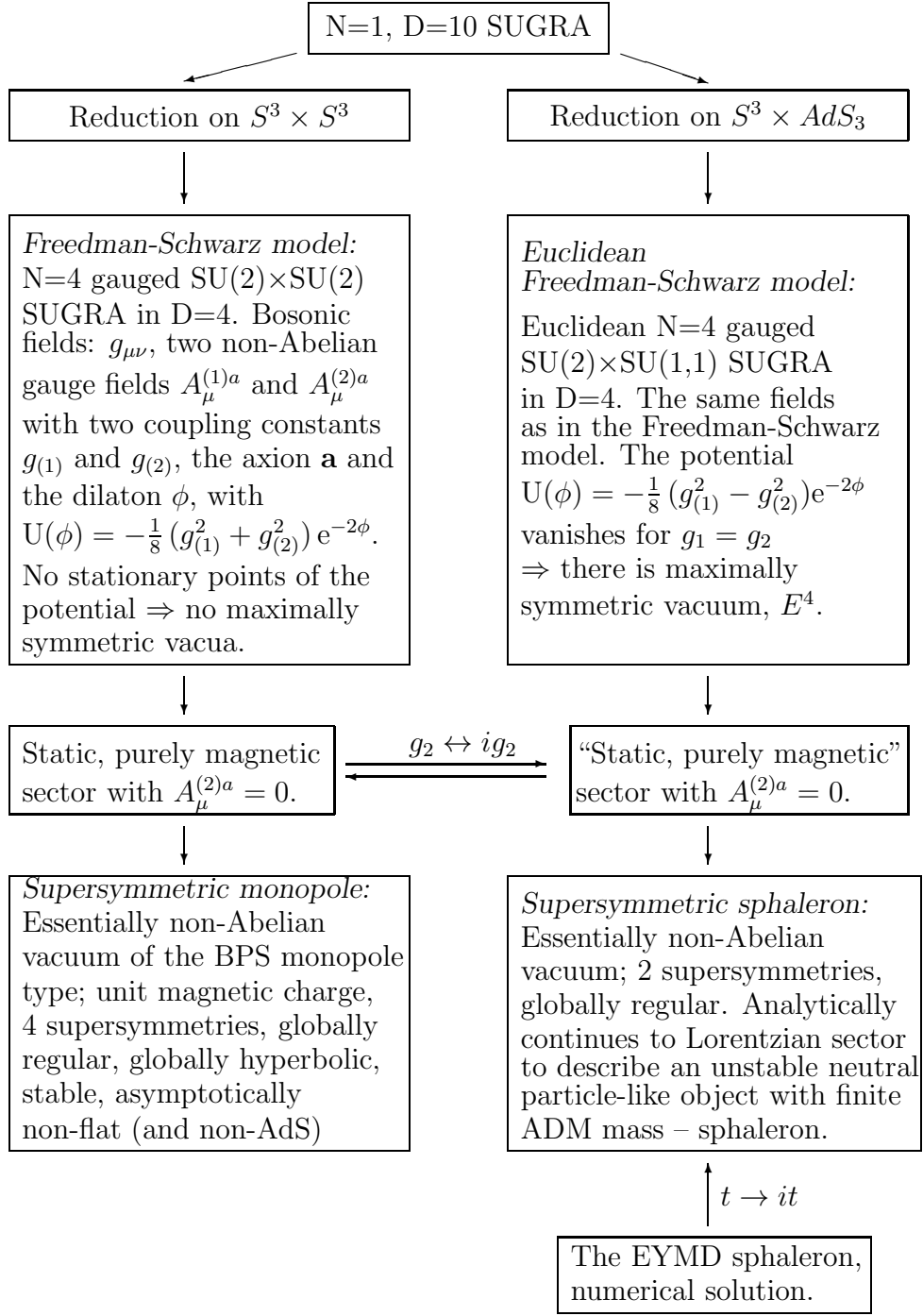
\begin{picture}(600,480)(-100,-350)

\put (20,115){\framebox(120,20){N=1, D=10 SUGRA}}
\put (15,115){\vector (-4,-1){45}}
\put (145,115){\vector (3,-1){35}}

\put (-100,80){\framebox(160,20){Reduction on $S^3\times S^3$}}
\put (100,80){\framebox(160,20){Reduction on $S^3\times AdS_3$}}

\put (180,75){\vector (0,-1){20}}
\put (-30,75){\vector (0,-1){20}}

\put (-100,-130){\framebox(160,180){}}
\put (-95,35){\shortstack{{\sl Freedman-Schwarz model:}}}
\put (-95,-125){\shortstack[l]{N=4 gauged 
SU(2)$\times$SU(2)\\SUGRA in D=4.
Bosonic \\fields: $g_{\mu\nu}$, two non-Abelian \\
gauge fields $A^{(1)a}_\mu$  and 
$A^{(2)a}_\mu$  \\ 
with two coupling 
constants \\$g_{(1)}$ and $g_{(2)}$, the axion ${\bf a}$ 
and \\the dilaton $\phi$, with \\
U$(\phi)=-\frac18\,(g_{(1)}^2+g_{(2)}^2)\,{\rm e}^{-2\phi}$.\\
No stationary points of the \\
potential $\Rightarrow$ no maximally \\
symmetric vacua.}}

\put (100,-130){\framebox(160,180){}}
\put (105,20){\shortstack[l]{{\sl Euclidean} \\ {\sl 
Freedman-Schwarz model:}}}
\put (105,-110){\shortstack[l]{Euclidean N=4 gauged \\
SU(2)$\times$SU(1,1) SUGRA \\
in D=4. The same fields \\
as in the Freedman-Schwarz \\model. The potential \\
U$(\phi)=-\frac18\,(g_{(1)}^2-g_{(2)}^2){\rm e}^{-2\phi}$\\
vanishes for $g_1=g_2$ \\ $\Rightarrow$ 
there is maximally \\
symmetric  vacuum, $E^4$.}}

\put (180,-135){\vector (0,-1){15}}
\put (-30,-135){\vector (0,-1){15}}

\put (-100,-190){\framebox(130,35){}}
\put (-95,-185){\shortstack[l]{
Static, purely magnetic \\sector with $A^{(2)a}_\mu=0$.}}

\put (125,-190){\framebox(135,35){}}
\put (127,-185){\shortstack[l]{
``Static, purely magnetic'' \\sector with $A^{(2)a}_\mu=0$.}}

\put (180,-195){\vector (0,-1){15}}
\put (-30,-195){\vector (0,-1){15}}

\put (60,-165){\shortstack[l]{$g_2\leftrightarrow ig_2$}}

\put (35,-170){\vector (1,0){80}}
\put (115,-175){\vector (-1,0){80}}

\put (-100,-325){\framebox(160,110){}}
\put (-95,-320){\shortstack[l]{{\sl Supersymmetric monopole:}\\
Essentially non-Abelian \\
vacuum of the BPS monopole \\
type; unit magnetic charge, \\
4 supersymmetries, 
globally \\ regular, 
globally hyperbolic, \\
stable, asymptotically \\
non-flat (and non-AdS) }}

\put (100,-325){\framebox(165,110){}}
\put (105,-320){\shortstack[l]{{\sl Supersymmetric sphaleron:}\\
Essentially non-Abelian \\
vacuum;
2 supersymmetries, \\
globally regular. 
Analytically \\continues to Lorentzian
sector \\ to describe an unstable neutral\\
particle-like object with finite \\ ADM mass -- 
sphaleron.}}

\put (120,-390){\framebox(125,30){}}
\put (125,-385){\shortstack[l]{The EYMD sphaleron,\\
numerical solution.}}

\put (180,-355){\vector (0,1){25}}
\put (185,-347){\shortstack[l]{$t\to it$}}
   
\end{picture}

\

\vspace{1 cm}
\noindent
{\bf Dimensional reduction of type I string theory.--}
Let us sketch the essential steps leading to the above diagram.
The details can be found in \cite{Volkov99a}, and also in 
\cite{Chamseddine97,Chamseddine98,Volkov99}. 
The starting point is the bosonic part of the action of D=10, N=1 supergravity
\begin{equation}
S_{10}=\int \left(
\frac{1}{4}\,\hat{R}
-\frac{1}{2}\,\partial_{M}\hat{\phi}\,\partial ^{M}\hat{\phi}
-\frac{1}{12}\,e^{-2\hat{\phi}}\,
\hat{H}_{MNP}\,\hat{H}^{MNP}\right)
\sqrt{-\hat{g}}\,d^{10}\hat{x}.                          \label{c0}
\end{equation}
The idea  is to find a parameterization of $\hat{g}_{MN}$, 
$\hat{H}_{MNP}$ and $\hat{\phi}$ in terms of 
four-dimensional variables which reduces the equations of motion
for the action (\ref{c0}) to a consistent system of 
four-dimensional equations.  As a first step, 
the tangent space metric is chosen in the form
\be                                                             \label{c6a}
\hat{\eta}_{AB}={\rm diag}
(\overbrace{\underbrace{s^2,+1,+1,+1}_{\rm 4-space}}^{\eta_{\alpha\beta}}
,\underbrace{\overbrace{+1,+1,+1}^{\eta^{(1)}_{ab}}
,\overbrace{+1,+1,-s^2}^{\eta^{(2)}_{ab}}}_{6-{\rm space}}).
\ee
Here the parameter $s$ assumes two values: $s=1$ or $s=i=\sqrt{-1}$.
The two options correspond
to the same theory in D=10 -- up to a renumbering of coordinates --
but to two different choices of the four-space. 
Keeping this parameter in all formulas, we shall be able 
to consider the two different cases together -- the 
reduction on $S^3\times S^3$ for $s=i$ and that 
on $S^3\times AdS_3$ for $s=1$.

The  D=10 metric is expressed in terms of the 
vielbein 
\be                                                     \label{c16}
\hat{{g}}^{MN}
\frac{\partial}{\partial\hat{x}^M}
\frac{\partial}{\partial\hat{x}^N}=\hat{\eta}^{AB}
\hat{E}_{A}\hat{E}_{B}=
\eta^{\alpha\beta}\hat{E}_{\alpha}\hat{E}_{\beta}
+\sum_{(\sigma)=1,2}
\eta^{(\sigma)ab}\hat{E}^{(\sigma)}_{a}\hat{E}^{(\sigma)}_{b},
\ee
where the vielbein vectors $\hat{E}_B=(\hat{E}_\alpha, \hat{E}^{(\sigma)}_a)$ 
are
\be                                                    
\hat{E}_\alpha={\rm e}^{3\phi/4}
\left(e_\alpha+\sum_{(\sigma)=1,2}e_\alpha^{\ \mu}
A^{(\sigma)a}_{\mu}\tilde{e}^{(\sigma)}_a\right),\ \ \ \
\hat{E}^{(\sigma)}_a
=-\frac{g_{(\sigma)}}{\sqrt{2}}\,
{\rm e}^{-\phi/4}\,\tilde{e}^{(\sigma)}_{a} . \label{c17}
\ee
Here  the tetrad four-vectors  
$e_\alpha\equiv e_\alpha^{\ \mu}\,\partial/\partial x^\mu$, 
the four-dilaton $\phi=-\frac12\,\hat{\phi}$,
and the fields $A^{(\sigma)a}_{\mu}$ depend only on the four-coordinates
$x^\mu$, the four-metric being
$g^{\mu\nu}=\eta^{\alpha\beta}e_\alpha^{\ \mu}e_\beta^{\ \nu}$.
Vectors 
$\tilde{e}^{(\sigma)}_{a}\equiv
\tilde{e}^{(\sigma)i}_{a}\partial/\partial z^{(\sigma)i}$
depend only on the internal six-coordinates $z$ and are assumed
to be invariant vectors of the internal group space ${\cal G}$.
Their commutation relations are
$[\tilde{e}_{a}^{(\sigma)},\tilde{e}_{b}^{(\sigma')}]=
\delta_{\sigma\sigma'}
f_{\ \ \ \ \, ab}^{(\sigma) c}\,\tilde{e}^{(\sigma)}_{c}$
with constant $f_{\ \ \ \ \, ab}^{(\sigma) c}$. 
It is assumed that the internal space  is a direct product 
of two group manifolds labeled by $(\sigma)=1,2$:
${\cal G}$=SU(2)$\times$SU(2) for $s=i$ and  
${\cal G}$=SU(2)$\times$SU(1,1) for $s=1$.
The structure constants for the two group factors are
$f_{\ \ \ \ ab}^{(\sigma)c}=\eta^{(\sigma)cd}\,\varepsilon_{dab}$.
The quantities $A^{(\sigma) a}_{\mu}$ are regarded as the 
non-Abelian gauge fields with the field strength
\be                                                       \label{c13}
F^{(\sigma) a}_{\mu\nu}=
\partial_{\mu}A^{(\sigma) a}_{\nu}-\partial_{\nu}A^{(\sigma) a}_{\mu}
+f_{\ \ \ \ bc}^{(\sigma) a}A^{(\sigma) b}_{\mu}A^{(\sigma) c}_{\nu}\, ,
\ee
with $g_{(\sigma)}$ in (\ref{c17})
 being the corresponding real gauge coupling constants. 

The ten-dimensional  antisymmetric tensor $\hat{H}_{MNP}$
has the following 
non-vanishing components  in the basis (\ref{c17})
(the value of $(\sigma)$ is the same for all internal indices):
\bea
\hat{H}^{(\sigma)}_{abc}&=&\frac{g_{(\sigma)}}{2\sqrt{2}}\,
{\rm e}^{-3\phi/4}\,\varepsilon_{abc}\, ,              \nonumber     \\
\hat{H}^{(\sigma)}_{a\alpha\beta}&=&-\frac{1}{\sqrt{2}g_{(\sigma)}}\,
{\rm e}^{5\phi/4}\,
\eta^{(\sigma)}_{ab}F^{(\sigma)b}_{\ \alpha\beta}\, ,\nonumber    \\
\hat{H}_{\alpha\beta\gamma}&=&{\rm e}^{-7\phi/4}\,
\varepsilon_{\alpha\beta\gamma}^{\ \ \ \ \,\delta}\,
e_\delta^{\ \mu}\partial_\mu{\bf a}\, .   \label{c18}
\eea
Here the axion ${\bf a}$ depends only on $x^\mu$, 
one has $\varepsilon^{0123}=1$, and 
$F^{(\sigma)b}_{\ \ \ \alpha\beta}$ are components of the gauge
field strength with respect to the tetrad $e_\alpha$.

As a result, all D=10 fields are expressed in terms of the four-dimensional
metric $g_{\mu\nu}$, two non-Abelian gauge fields $A^{(\sigma) a}_{\mu}$,
the axion ${\bf a}$ and the dilaton $\phi$.  These expressions are
then inserted into the D=10 field equations for the action (\ref{c0}).
It turns out that all these equations are fulfilled provided that the 
four-dimensional fields satisfy the 
 equations derived from the following Lagrangian:
\bea
{\cal L}_4&=&\frac{R}{4}\,-\frac12\,\partial_\mu\phi \,\partial^\mu\phi
+\frac{s^2}{2}\,{\rm e}^{-4\phi}\, 
\partial_\mu{\bf a}\,\partial^\mu{\bf a}  
-\frac14\,
{\rm e}^{2\phi}\sum_{(\sigma)=1,2}\frac{1}{g_{(\sigma)}^2}\,
\eta^{(\sigma)}_{ab}
F^{(\sigma)a}_{\ \,\mu\nu}F^{(\sigma)b \mu\nu}  \nonumber  \\
&-&\frac{1}{2}\,{\bf a}
\sum_{(\sigma)=1,2}\frac{1}{g_{(\sigma)}^2}\,\eta^{(\sigma)}_{ab}
\ast\! F^{(\sigma)a}_{\ \,\mu\nu}F^{(\sigma)b\mu\nu}
+\left.\left.\frac18\,                                  
\right(g_{(1)}^2-s^2g_{(2)}^2\right){\rm e}^{-2\phi}  \, . \label{c26}   
\eea
We thus obtain the four-dimensional theory upon  a consistent
dimensional reduction from D=10. 
Let us recall that  the signature
of the spacetime metric
is $(s^2,+1,+1,+1)$,
the internal metrics are $\eta^{(1)}_{ab}=\delta_{ab}$ and 
$\eta^{(2)}_{ab}={\rm diag}(+1,+1,-s^2)$. 
For $s=i$ the theory (\ref{c26}) exactly coincides with the bosonic
sector of the Freedman-Schwarz model \cite{Freedman78},
which is the N=4 gauged SU(2)$\times$SU(2) supergravity.   
For $s=1$ we obtain the new theory -- Euclidean 
N=4 gauged SU(2)$\times$SU(1,1) supergravity.
Notice that in the latter case the dilaton potential is proportional
to the difference and not the sum of the two coupling constants.

The bosonic Lagrangian (\ref{c26}) is to be supplemented by the 
rules for computing the fermionic SUSY variations. These can be
derived via the dimensional reduction of the D=10 SUSY variations 
\bea
\delta \hat{\psi }_{M}&=&\hat{\cal D}_M\,\hat{\epsilon}
-\frac{1}{48}\, e^{-\hat{\phi}}\,           
\left( \hat{\Gamma}_{\ \ \ \  M}^{SPQ}+9\,\delta _{M}^{S}\,
\hat{\Gamma}^{PQ}\right) \,\hat{H}_{SPQ}\,\hat{\epsilon}\, , \nonumber  \\
\delta \hat{\chi }&=&-\frac{1}{\sqrt{2}}\, 
(\hat{\Gamma}^{M}\partial _{M}\hat{\phi})\,\hat{\epsilon}
-\frac{1}{12\sqrt{2}}\,e^{-\hat{\phi}}%
 \,\hat{\Gamma}^{SPQ}\, \hat{H}_{SPQ}\,\hat{\epsilon}\, .  \label{c5}
\eea
Here $\hat{\epsilon}$ is the Majorana-Weyl spinor parameter  
of supersymmetry transformations, $\hat{\cal D}_{M}$ being 
its covariant derivative, and $\hat{\Gamma}^{A}$ are D=10 
gamma matrices. The idea  now is to choose an 
explicit parameterization of $\hat{\Gamma}^{A}$ in terms of D=4
gamma matrices $\gamma^\mu$ and the gauge group generators
$\T^{(\sigma)}_a$, where
$[\T^{(\sigma)}_a,\T^{(\sigma)}_b]=\varepsilon_{abc}\,\eta^{(\sigma)cd} \, 
\T^{(\sigma)}_d$.  Also 32-component Majorana-Weyl 
spinors  $\delta \hat{\psi }_M$,  $\delta \hat{\chi }$, 
and $\hat{\epsilon}$ are expressed in terms of the four-dimensional 
spinors
$\delta {\psi }_\mu$,  $\delta{\chi }$, 
and ${\epsilon}$; details can be found in \cite{Volkov99a}.
Here each four-dimensional spinor, 
${\epsilon }$ say, can be thought of as a multiplet
of D=4 Majorana spinors in the fundamental representation
of the gauge group ${\cal G}$. One can write 
$\epsilon\equiv\epsilon^{\rm I}_\kappa$, where I$=1,\ldots 4$ is 
the group index and  $\kappa=1,\dots 4$ is the spinor index.  
The important moment is the D=4 Majorana condition, which 
is derived from the D=10 Majorana-Weyl condition:
\be                                                     \label{d17}
\epsilon^\ast=A\otimes B\,\epsilon\, .
\ee
Here the asterisk denotes complex conjugation, and 
$4\times 4$ matrices $A$ and $B$ are such that
\be                                                     \label{d16}
A\gamma^\alpha A^{-1}=\left(\gamma^{\alpha}\right)^\ast,\ \ \
B\T^{(\sigma)}_{a} B^{-1}=\left(\T^{(\sigma)}_{a}\right)^\ast\, ,
\ee
and they fulfill also the condition of 
consistency of (\ref{d17}):
\be                                                     \label{d16a}
AA^\ast\otimes BB^\ast=\1\, .
\ee
It is worth emphasizing that the Majorana condition (\ref{d17})
 is consistent
also in the Euclidean case. This is only possible 
due to the group degrees of freedom: because the group 
matrix $B$ is present
in (\ref{d16}), (\ref{d16a}). Specifically, 
in the Euclidean case  each of the 
two factors in (\ref{d16a}) has wrong sign, 
$AA^\ast=-1$ and $BB^\ast=-1$, but their product has 
correct sign.

Omitting further details, the reduction of (\ref{c5}) gives the 
following SUSY variation for the D=4 gaugino:
\bea                                            \label{d24}
\delta\chi&=&\left(\frac{1}{\sqrt{2}}\,\gamma^\mu\partial_\mu\phi-
\frac{1}{\sqrt{2}s}\,{\rm e}^{-2\phi}
\gamma_5\gamma^\mu\partial_\mu{\bf a}\right)
\epsilon \nonumber \\
&+&\frac{1}{2s}\,{\rm e}^\phi
\left(s{\cal F}^{(1)}-\gamma_5{\cal F}^{(2)}\right)
\epsilon
+\frac{1}{4s}\,{\rm e}^{-\phi} \left(s\,g_{(1)}-
g_{(2)}\gamma_5\right)\epsilon\, ,
\eea
and for the gravitino:
\bea                                                    \label{d32}
&&\delta\psi_\mu=\left(\partial_\mu
+\frac14\,\omega_{\alpha\beta,\mu}\gamma^\alpha\gamma^\beta
+\,\sum_{(\sigma)=1,2}K^{(\sigma)}_{ab}
\T^{(\sigma)a}A_{\mu}^{(\sigma)b}+
\frac{1}{2s}\,{\rm e}^{-2\phi}\gamma_5\,\partial_\mu{\bf a}\right)
\epsilon \nonumber \\
&&+\frac{1}{2\sqrt{2}s}\,{\rm e}^\phi
\left(s{\cal F}^{(1)}+\gamma_5{\cal F}^{(2)}\right)
\gamma_\mu\epsilon
+\frac{1}{4\sqrt{2}s}\,{\rm e}^{-\phi} 
\left(s\,g_{(1)}+g_{(2)}\gamma_5\right)\gamma_\mu\epsilon\, .
\eea
Here ${\cal F}^{(\sigma)}=-\frac{1}{g_{(\sigma)}}\,
\eta^{(\sigma)}_{ab}\,\gamma^\alpha\gamma^\beta\,
F_{\alpha\beta}^{(\sigma)a}\T^{(\sigma)b}$ and 
$K^{(1)}_{ab}=\eta^{(1)}_{ab}$, 
$K^{(2)}_{ab}=-s^2\eta^{(2)}_{ab}$ with
$\T^{(\sigma)}_a=\T^{(\sigma)a}$.
These expressions complete the dimensional reduction procedure. 
For $s=i$ Eqs.(\ref{d24}) and (\ref{d32}) exactly coincide with the 
linearized fermionic SUSY variations in the Freedman-Schwarz model
\cite{Freedman78}, while giving for $s=1$ the rules for the
Euclidean Freedman-Schwarz model. 
It is important that the reduction is consistent in the sense that 
if the four-dimensional configuration is on-shell, then its
uplifted version is a solution of the D=10 equations. 
Also, if the D=4 SUSY variations vanish, then the D=10 variations
vanish too.

The FS and EFS models are related to each other via
$s\leftrightarrow is$ and not via $g_{(2)}\leftrightarrow ig_{(2)}$. 
However, both models can be
consistently truncated to the static, purely magnetic sector by 
setting $A^{(2)a}_\mu=A^{(1)a}_0={\bf a}=0$ and requiring 
$\frac{\partial}{\partial x^0}$ to be a hypersurface orthogonal
Killing vector. The direct inspection of the field
equations for the bosonic Lagrangian (\ref{c26}) and the SUSY
variations (\ref{d24}) and (\ref{d32}) reveals then that 
$s\leftrightarrow is$ becomes equivalent to 
$g_{(2)}\leftrightarrow ig_{(2)}$. This explains
the empirical relation in (\ref{0a}). 

\noindent
{\bf Bogomol'nyi equations.--} The D=4 models obtained above
admit  supersymmetric vacua. 
These are solutions of the 
bosonic field equations for which there are such non-trivial spinors
$\epsilon$ that $\delta\chi=\delta\psi_\mu=0$. 
It is rather difficult to directly solve the second order field equations for 
the 
bosonic Lagrangian (\ref{c26}), especially with non-trivial gauge fields. 
The procedure is therefore to focus on the conditions 
$\delta\chi=\delta\psi_\mu=0$, and this gives
the system of  equations for the $\epsilon$. 
These equations are usually inconsistent, but one can look for the
consistency conditions. The latter can be formulated as a system of 
first order equations for the bosonic background, usually called
Bogomol'nyi equations.  The Bogomol'nyi equations are compatible with 
the second order field equations, and are sometimes integrable.

To find the Bogomol'nyi equation we first truncate the system to the 
static, purely magnetic case as described above, and then impose
in addition the spherical symmetry:
\be                                                 \label{a2}
ds^2=s^2{\rm e}^{2(\phi-\phi_\infty)}dt^2+
{\rm e}^{2\lambda}\,(dr^2+
r^2\,d\Omega^2), 
\ee
\be                                                     \label{aa2}
A\equiv A^{(1)}=w\ (-\T_2\,d\theta +\T_1\,
\sin \theta \,d\varphi )+\T_3\,\cos \theta \,d\varphi .
\ee
Here $\phi$, $\lambda$ and $w$ are functions of the radial coordinate $r$,
and one can show that the ``metric-dilaton relation'' 
$g_{00}=s^2{\rm e}^{2(\phi-\phi_\infty)}$ is consistent
with the field equations.
The procedure is then to insert (\ref{a2}), (\ref{aa2}) 
into (\ref{d24}) and (\ref{d32}) 
and set the left-hand sides to zero, which gives a system of equations 
for $\epsilon$. The next step is to restrict $\epsilon$ to the sector
with zero total (orbital+spin+isospin) angular momentum. 
The equations then essentially become algebraic, in which case 
the consistency conditions
can be easily obtained \cite{Volkov99a}. This gives the 
Bogomol'nyi equations
\bea
1+r\frac{d\lambda}{dr}&=&
\sqrt{w^2+\frac{1}{8}\,
 {\rm e}^{2(\lambda+\ln r-\phi)}\,
\left((B-1)^2-\xi^2\right)}\equiv\nu \, ,  \nonumber
\\
A\,r\frac{dw}{dr}&=&
2\xi\,w\nu+\xi^2(w^2-1)-2w^2(B+1)\, , \nonumber
\\
A\,r\frac{d\phi}{dr}&=&-(B+1)
(\xi\nu+w(B-1))\, .                  \label{a26}
\eea
Here
$A\equiv 8w\nu\,{\rm e}^{2(\phi-\lambda-\ln r)}
+\xi\,(B-1)$ and $B\equiv 2{\rm e}^{2(\phi-\lambda-\ln r)}(w^2-1)$,
and also $\xi=g_{(2)}/s$. These equations are compatible 
with the second order field equations for the Lagrangian (\ref{c26}), and  
for any solution there are two supersymmetry Killing spinors.
If $\xi =0$ then the number of
supersymmetries doubles. 

\noindent
{\bf Supersymmetric monopole.--}
Let us set in (\ref{a26}) $\xi=0$. Notice that the equations
then become the same both in the FS and EFS cases. 
After  the substitution \cite{Chamseddine97}
\be
w^2=\rho^{2}\,e^{y(\rho)},\ \ \ \ \ \ \
\frac{1}{2}\,e^{2(\lambda+\ln r-\phi) }
=-\rho\frac{dy (\rho)}{d\rho}-\rho^{2}\,e^{y
(\rho)}-1,                                             \label{a28}
\ee
the Bogomol'nyi equation (\ref{a26}) become equivalent 
to one equation
\be
\frac{d^{2}y}{d\rho^{2}}=2\, e^{y},  \label{a29}
\ee
which is integrable. This gives the solution in the closed form:
\be
d{ s}^{2}=2\,{\rm e}^{2\phi}\left(
s^2dt^{2}+d\rho^{2}+R\,d\Omega^2\right) ,  \label{a30}
\ee
\be                                                   \label{a31}
R=2\rho\coth \rho-\frac{\rho^{2}}{\sinh ^{2}\rho}-1,\ \ 
w=\pm \frac{\rho}{\sinh \rho},\ \ 
{\rm e}^{2\phi }= \frac{\sinh \rho}{\sqrt{R}}.    
\ee
For $s=1$ (the EFS case) this solution describes a globally Euclidean
background with an essentially non-Abelian gauge field.
Since the configuration does not depend on $t$, the action
is infinite. For $s=i$ (the FS case)
the solution  is of the BPS monopole
type with unit magnetic charge. This is geodesically complete and 
globally hyperbolic \cite{Chamseddine97,Chamseddine98}. 
Unfortunately, the ADM mass is infinite and the solution
is not asymptotically flat -- due to the dilaton potential. 
In view of its supersymmetry, it is very plausible that the 
Lorentzian solution is stable, while for its Euclidean 
counterpart the notion of dynamical stability makes no sense.

\noindent
{\bf Supersymmetric sphaleron.--}
Let us set in (\ref{a26}) $\xi=1$, in which case the theory is 
Euclidean and the dilaton potential vanishes. 
After some transformations 
described in \cite{Volkov99}
the Bogomol'nyi equations can be reduced to one:
\be                                              \label{a33}
\frac{1}{2r}\,\frac{dw}{dr}=\frac{1-w^2}{4r^2}
-\frac{(w+1)^3}{8}+\frac{(w-1)^3}{8r^4}\, ,
\ee
which is invariant under $r\to1/r$, $w\to-w$.
When  the solution $w(r)$ is found, the metric function 
$\lambda$ is obtained from 
\be                                                   \label{a35}
\lambda=
\ln(2)+\int_{0}^{r}\left(\frac{U+w}{2}-1\right)\frac{dr}{r},\ \ \ 
{\rm with}\ \ \ \ 
U=\frac{r^2(1+w)^2+(1-w)^2}{r^2(1+w)^2-(1-w)^2}\, ,
\ee
while the dilaton is given by 
\be                                                  \label{a36}
\phi=
\lambda+\ln(r)+\frac12\,
\ln\left(\frac{(U+w)^2-2w^2-2}{2\,(w^2-1)^2}\right)\, ,
\ee
which is normalized such that $\phi(0)=0$. 
Unfortunately, analytical solutions to Eq.(\ref{a33}) are unknown.
The numerical integration
reveals the existence of a globally regular solution
in the interval $r\in[0,\infty)$ which monotonically interpolates 
between the values specified by the local asymptotic solutions:
$w=1-\frac23\,r^2+O(r^4)$ for $r\to 0$ and 
$w=-1+2\sqrt{2}\,r^{-1}+O(r^{-2})$ as $r\to\infty$. 
This gives a globally regular 
supersymmetric Euclidean solution with  non-trivial
Yang-Mills field and infinite action.

Now, one can pass to the Lorentzian sector 
by simply changing the sign of  $dt^2$  in the metric (\ref{a2}).
Of course, this will not lead to the Freedman-Schwarz
theory (we do not replace $s\to is$
but just $t\to it$), but rather to a solution 
of the Einstein-Yang-Mills-Dilaton  model. This 
describes an unstable regular
particle-like object with finite ADM mass 
and purely magnetic Yang-Mills amplitude -- the EYMD sphaleron.
Despite its instability, this solution still bears 
an imprint of supersymmetry
as it fulfills the same system of first order Bogomol'nyi equations
as its Euclidean counterpart.  Passing
back to the Euclidean theory, the full supersymmetry is restored,
and we call the Euclidean solution
``supersymmetric sphaleron''. 

The above considerations hopefully clarify somewhat the tale of the 
EYMD sphaleron.
The story, however, is not yet finished, as the Bogomol'nyi
equations (\ref{a33}) is still not solved analytically. 
Also it is unclear what
happens when the Euclidean solution is uplifted to D=10. 
In the limits $r\to 0,\infty$ the geometry of the solution is flat
and the gauge field vanishes, in which case the uplifted solution is 
$E^4\times S^3\times AdS_3$. This is known to describe the 
near-horizon geometry of a system of parallel  D5--D1 branes.
If we consider the full range of $r$, then the uplifted solution 
presumably describes the near-horizon geometry of some
deformation of the D5--D1 system.
It is interesting to know what this configuration is. 
Finally, the EYMD sphaleron itself
can also be lifted to ten dimensions as
a non-supersymmetric solution of heterotic string theory.  
It is then interesting to know whether there is any
relation to the recently discussed ``D-sphalerons'' \cite{Harvey00}.

\end{document}